\documentclass[11pt,preprint]{aastex} 

\usepackage{hyperref}
\usepackage{natbib}
\received{}
\accepted{}
\shorttitle{Cepheids in M31}
\shortauthors{Riess et al.}

\newcommand{\bq}{\begin{equation}} 
\newcommand{\eq}{\end{equation}}

\newcommand{\beq}{\begin{equation}}
\newcommand{\eeq}{\end{equation}}
\newcommand{\beqa}{\begin{eqnarray}}
\newcommand{\eeqa}{\end{eqnarray}}

\newcommand{\PL}{$P$--$L$\ }

\def\deg{\hbox{$^\circ$}}
\def\sun{\hbox{$\odot$}}

\def\arcsec{\hbox{$^{\prime\prime}$}}

\def\farcsec{\hbox{$.\!\!^{\prime\prime}$}}

\begin{document} 

\title{Cepheid Period-Luminosity Relations in the Near-Infrared \\
and the Distance to M31\\
from the Hubble Space Telescope Wide Field Camera 3
\altaffilmark{*}}

\author{Adam G. Riess\altaffilmark{1,2}, J\"urgen Fliri\altaffilmark{3} and
David Valls-Gabaud\altaffilmark{3}}

\altaffiltext{*}{Based on observations with the NASA/ESA {\it Hubble Space
  Telescope}, obtained at the Space Telescope Science Institute, which is
  operated by AURA, Inc., under NASA contract NAS 5-26555.}

\altaffiltext{1}{Department of Physics and Astronomy, Johns Hopkins
  University, Baltimore, MD 21218.}
\altaffiltext{2}{Space Telescope Science Institute, 3700 San Martin
  Drive, Baltimore, MD 21218; ariess@stsci.edu}
\altaffiltext{3}{LERMA, CNRS UMR 8112, Observatoire de Paris,
  61 Avenue de l'Observatoire, 75014 Paris, France; 
jurgen.fliri@obspm.fr, david.valls-gabaud@obspm.fr}

\begin{abstract} 
  We present measurements of 68 classical Cepheids with periods from 10 to 
78 days observed in the near-infrared by the PHAT Program using the Wide 
Field Camera 3 (WFC3) on the {\it Hubble Space Telescope
  (HST)}.   The combination of HST's resolution and the use of near-infrared 
measurements provides a dramatic reduction in the dispersion of the 
Period--Luminosity 
relation over the present optical, ground-based data.  Even using random-phase 
magnitudes we measure a dispersion of just 0.17 mag, implying a 
dispersion of just 0.12 mag for mean magnitudes.  The error in 
the mean for this relation is 1\% in distance.  Combined with similar 
observations of Cepheids in other hosts and independent distance 
determinations, we measure a distance to M31 of 
$\mu_0=24.42 \pm 0.05 (statistical) \pm 0.03 (systematic)$, 765 $\pm$ 28 kpc,
in good agreement with past measurements though with a better, 3\% precision here.  
The result is also in good agreement with independent distance 
determinations from two detached eclipsing binaries allowing for an 
independent calibration of the Cepheid luminosities and a determination
 of the Hubble constant.
  \end{abstract} 


\section{Introduction} 
M31, the nearest analogue of the Milky Way Galaxy, has long provided important 
clues to understanding the scale of the Universe.  The naked-eye visibility in 
1885 of a supernova in M31 \citep[S And, see][]{1985ApJ...295..287D} 
once suggested the spiral nebulae were located within the Milky Way lest
 its luminosity be ``\textit{on a scale of magnitude such as the imagination 
recoils from contemplating}'' \citep{clerke1903problems}. 
\citeauthor{1929ApJ....69..103H}'s \citeyearpar{1929ApJ....69..103H} 
subsequent discovery of Cepheid variables in M31 revealed the true gulf that 
existed between the Milky Way and other galaxies.  The ability to resolve the 
stellar populations of M31 and knowledge of its distance still provides the best 
constraints on the timescales of the formation of such massive galaxies 
\citep{2006ApJ...636L..89B}. 
Despite long past success in identifying Cepheids in M31 
\citep{1963AJ.....68..435B,1965AJ.....70..212B}, 
comprehensive inventories of its variables were not completed until the 
availability of wide-format CCD arrays to survey its 3 degree span and 
image-subtraction techniques to contend with the extreme crowding of its stars, 
a consequence of its 77\deg~inclination.  Large-scale variability  
 surveys in the 1990s  by 
\citet{1997A&AS..126..401M}, 
the DIRECT Program  \citep[e.g.][]{2003AJ....126..175B}, 
and \cite{2007A&A...473..847V},  
as well as microlensing surveys in the 2000s, by 
POINT-AGAPE \citep{2004MNRAS.351.1071A} 
and WeCAPP \citep{2006A&A...445..423F}, 
succeeded in discovering $\sim$ 10$^3$ Cepheids, most from $>$ 50 Myr-old 
supergiants at short periods ($P < 10$ days).  
The Pixel Observations of M31 with MEgacam (POMME) survey has been the most 
prolific to date, making use of Megacam on CFHT to discover more than 2500 
Cepheids \citep{fliri2012}.

  The utility of Cepheids as distance indicators critically depends on the 
accuracy of their measured fluxes.  Even the best optical measurements of M31 Cepheids 
from the ground have been biased bright at the $0.1-0.2$ mag level by stellar 
crowding \citep{2000AJ....120..810M,2007A&A...473..847V} 
 and contaminated by variable extinction owing to the high inclination of its host.  
  Space-based observations in the near-infrared have the ability to greatly 
mitigate both of these sources of error.  A modest sample of 8 Cepheids in M31
 was measured with greater resolution and in the near-infrared using the NICMOS 
Camera on the Hubble Space Telescope (HST) by \citet{2001ApJ...549..721M}, 
suggesting that both crowding and extinction could be tamed by such measurements.

  The Panchromatic Hubble Andromeda Treasury (PHAT) Program (P.I. J. Dalcanton) is 
a Hubble Space Telescope Multi-cycle Treasury program to map roughly a third of 
M31's star forming disk, using 6 filters covering from the ultraviolet through 
the near-infrared. With HST's resolution and sensitivity, the disk of M31 will 
be resolved into more than 100 million stars, enabling a wide range of scientific 
endeavors.  Despite capturing only a static view of M31, the random phase 
observations of known Cepheids obtained by PHAT are nearly as precise as mean 
phase observations for measuring the distance to M31 due to the small amplitudes
 of Cepheid light curves in the near-infrared \citep{1991PASP..103..933M}. 
Because the Cepheid observations are obtained with the same near-infrared photometric system as
 recent distance scale data \citep{2011ApJ...730..119R}, 
the PHAT observations provide the means to determine the distance to M31 with 
lower systematic error than past estimates.  The recent discovery and 
characterization of detached eclipsing binary sytems (DEB) in M31 
\citep{2005ApJ...635L..37R,2010A&A...509A..70V} 
 offers reliable and independent distance determinations to M31 which
 can be used to calibrate Cepheid luminosities and ultimately the Hubble 
constant.  These measurements are also valuable for characterizing the slope 
of the period-luminosity relation at solar metallicity at long-periods in the 
same near-infrared band used to determine the Hubble constant \citep{2011ApJ...730..119R}.
    
  We have analyzed the first year of PHAT data to locate and measure the 
long-period Cepheids ($\log P({\mathrm{day}}) >$ 1)  previously discovered by image-subtraction 
in ground-based variability surveys, mostly from POMME.  In Section 2 we present 
images of the recovered long-period Cepheids from the PHAT survey and their 
near-infrared photometry.  In Section 3 we analyze this data to constrain the 
distance to M31, the slope of the \PL relation in $F160W$ and its impact on 
the Hubble constant.

\section{WFC3 Observations of Cepheids in M31}

The PHAT program \citep{dalcanton2012} is imaging the northeast quadrant of M31 
with WFC3 with 2 UV filters ($F275W$, $F336W$), 2 IR filters ($F110W$, $F160W$) 
and 2 optical filters with ACS ($F475W$, $F814W$) over the course of a 3 year 
survey.  All WFC3 images are collected in a single epoch with those from ACS
 obtained approximately 6 months
before or after the WFC3 images when the orientation of HST allows overlapping 
coverage in parallel.  We searched the PHAT data for Cepheids with $\log P >$ 1
 as these match the period range of Cepheids observed by HST at $>20$ Mpc used 
in studies of the distance scale.  Our primary source of positions and periods 
of Cepheids was the recent POMME Survey from Megacam on CFHT whose sample of 
$>$ 2500 Cepheids covering most of M31 is the largest sample collected to date 
\citep{fliri2012}, 
with 180 in the North half of M31 with $\log P >$ 1.  While many of the Cepheids 
found in previous surveys are included in the POMME sample, the reverse is not 
true.  We also consulted the variability catalog from the DIRECT program 
\citep{2003AJ....126..175B} 
to search for additional long-period Cepheids not included in the POMME sample. 
 In the first year PHAT  obtained data through the middle of 2011, and 65 
long-period POMME Cepheids were contained in the survey, a fair fraction of the $\sim$ 90 
likely to be imaged by the end of the Survey.  These 65 were augmented with 1 
object from DIRECT and 2 from the Pan-STARRS survey data \citep{2011arXiv1109.6320L}. 

\begin{figure}[ht]
\vspace*{140mm}
\figurenum{1}
\includegraphics{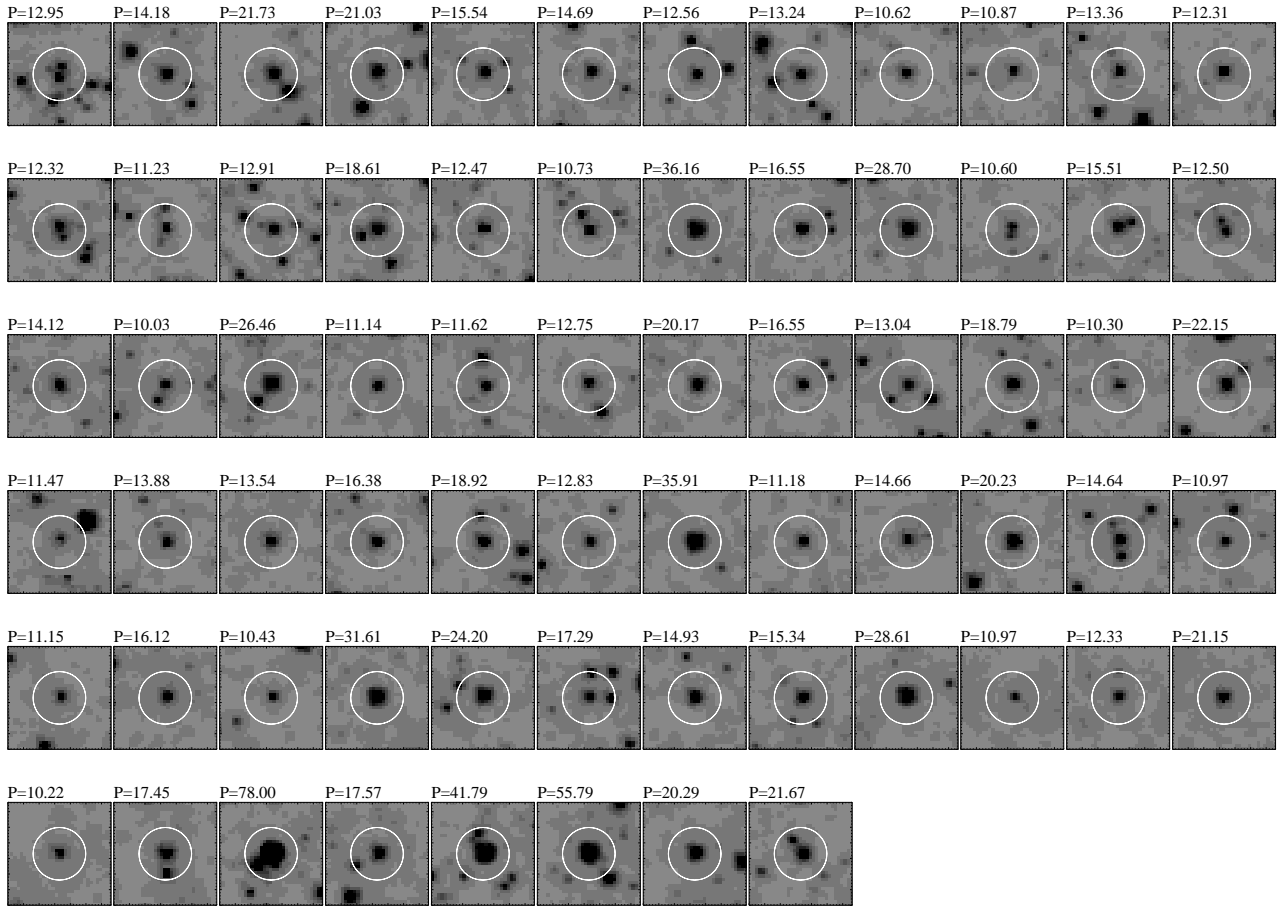}
\caption{HST near-IR $F160W$ images of 68 Cepheids in M31
(ordered as in Table~\ref{table1}), spanning nearly an order of magnitude in period.
The scale of each stamp is 2.5\arcsec (32 pixels). The position of the Cepheid
as determined from the optical Megacam POMME images in indicated by the
circle, which has a diameter of 1\arcsec. \label{fig1}}
\end{figure}

Precise positions of the Cepheids in the HST images of M31 were determined by 
refining steps in the relative astrometry between ground- and space-based imaging. 
 The ground-based survey positions were first used to identify the approximate 
region hosting the Cepheids in the HST images to $\sim$ 0\farcsec4  precision, 
with the uncertainty resulting from the HST guide star position errors.  Next, 
a geometric transformation was defined by matching unresolved sources in
 20\arcsec~diameter regions
in the ground-based $i$-band POMME data and the WFC3 $F110W$ images.  The derived 
transformations were used with the POMME Cepheid coordinate to locate the Cepheid 
positions to within one WFC3-IR pixel ($\sigma$$\sim$0\farcsec1).  For a few Cepheids 
with close neighbors, identifications of the Cepheids in the $F336W$ images were 
used to confirm the Cepheid position among neighboring red giants (missing in 
the UV) or luminous blue dwarves (missing in the IR).  Lastly, centroids for the
 Cepheids in the HST images
were measured together with photometry resulting in the positions given in 
Table~\ref{table1}.  The high signal-to-noise ratio of the Cepheid data in the HST images 
(SNR $>$ 100) and minimal crowding (see Figure~\ref{fig1}) insures negligible bias in the 
measured positions.    

\clearpage
\tabletypesize{\scriptsize}
\tablewidth{0pt}
\begin{deluxetable}{llllllll}
\tablenum{1}
\tablecaption{M31 Cepheids in WFC3-IR \label{table1}}
\tablehead{\colhead{$\alpha$}&\colhead{$\delta$}&
\colhead{Id\tablenotemark{a}}&\colhead{P}&\colhead{$F160W^b$}&\colhead{$F110W$}&\colhead{Bias}&\colhead{[O/H]} \\ 
\colhead{(J2000)}&\colhead{(J2000)}&\colhead{ }&\colhead{(day)}&\colhead{(mag)}&\colhead{(mag)}&\colhead{(mag)}& \colhead{zkh}} 
\startdata
 10.896242  &  41.258875  & vn.5.1.1120 &  12.95  &  18.78 ( 0.05 ) & 19.09 ( 0.09 ) & 0.009  &  9.008  \\ 
 11.202987  &  41.487450  & vn.3.1.518 &  14.18  &  18.25 ( 0.03 ) & 18.42 ( 0.06 ) & 0.003  &  8.948  \\ 
 11.235392  &  41.472761  & vn.3.1.357 &  21.73  &  18.11 ( 0.03 ) & 18.62 ( 0.07 ) & 0.002  &  8.922  \\ 
 11.200633  &  41.449286  & vn.3.1.535 &  21.03  &  18.05 ( 0.03 ) & 18.54 ( 0.06 ) & 0.003  &  8.930  \\ 
 11.209271  &  41.452631  & vn.3.1.484 &  15.54  &  18.46 ( 0.03 ) & 18.98 ( 0.08 ) & 0.004  &  8.927  \\ 
 11.182408  &  41.476428  & vn.3.1.663 &  14.69  &  18.53 ( 0.03 ) & 18.99 ( 0.08 ) & 0.002  &  8.954  \\ 
 11.185921  &  41.417044  & vn.3.1.622 &  12.56  &  18.67 ( 0.03 ) & 19.04 ( 0.08 ) & 0.002  &  8.922  \\ 
 11.131458  &  41.632081  & vn.4.2.246 &  13.24  &  18.55 ( 0.03 ) & 18.83 ( 0.07 ) & 0.004  &  9.028  \\ 
 11.138413  &  41.626025  & vn.4.2.197 &  10.62  &  18.54 ( 0.03 ) & 18.84 ( 0.07 ) & 0.008  &  9.025  \\ 
 11.091775  &  41.664158  & vn.4.2.508 &  10.87  &  18.87 ( 0.03 ) & 19.22 ( 0.09 ) & 0.004  &  9.033  \\ 
 11.038733  &  41.662008  & vn.4.2.849 &  13.36  &  18.61 ( 0.03 ) & 18.88 ( 0.08 ) & 0.008  &  9.034  \\ 
 11.039342  &  41.659111  & vn.4.2.845 &  12.31  &  18.50 ( 0.03 ) & 18.65 ( 0.07 ) & 0.012  &  9.035  \\ 
 11.129060  &  41.613420  & vn.4.2.268 &  12.32  &  18.45 ( 0.03 ) & 18.75 ( 0.07 ) & 0.004  &  9.027  \\ 
 11.046342  &  41.654658  & vn.4.2.781 &  11.23  &  18.87 ( 0.03 ) & 19.09 ( 0.08 ) & 0.000  &  9.037  \\ 
 11.096287  &  41.588286  & vn.4.2.480 &  12.91  &  18.48 ( 0.03 ) & 18.64 ( 0.07 ) & 0.002  &  9.035  \\ 
 10.953500  &  41.624042  & vn.5.2.475 &  18.61  &  17.95 ( 0.03 ) & 18.44 ( 0.06 ) & 0.004  &  9.035  \\ 
 10.965058  &  41.620431  & vn.5.2.390 &  12.47  &  18.46 ( 0.03 ) & 18.80 ( 0.07 ) & 0.008  &  9.040  \\ 
 10.972425  &  41.631158  & vn.5.2.333 &  10.73  &  18.71 ( 0.04 ) & 18.88 ( 0.08 ) & 0.004  &  9.036  \\ 
 11.348500  &  41.693375  & vn.2.2.497 &  36.16  &  17.34 ( 0.03 ) & 17.93 ( 0.05 ) & 0.001  &  8.952  \\ 
 11.354621  &  41.708811  & vn.2.2.463 &  16.55  &  18.62 ( 0.03 ) & 18.74 ( 0.07 ) & 0.003  &  8.953  \\ 
 11.363342  &  41.699800  & vn.2.2.408 &  28.70  &  17.62 ( 0.03 ) & 18.04 ( 0.05 ) & 0.001  &  8.946  \\ 
 11.368488  &  41.659503  & vn.2.2.378 &  10.60  &  19.15 ( 0.03 ) & 19.30 ( 0.09 ) & 0.008  &  8.930  \\ 
 11.377304  &  41.657669  & vn.2.2.340 &  15.51  &  18.24 ( 0.03 ) & 18.53 ( 0.06 ) & 0.002  &  8.925  \\ 
 11.326300  &  41.650592  & vn.2.2.647 &  12.50  &  19.00 ( 0.03 ) & 19.32 ( 0.09 ) & 0.003  &  8.948  \\ 
 11.331008  &  41.666139  & vn.2.2.616 &  14.12  &  18.34 ( 0.03 ) & 18.57 ( 0.06 ) & 0.002  &  8.951  \\ 
 11.276512  &  41.694817  & vn.3.2.299 &  10.03  &  18.79 ( 0.03 ) & 19.06 ( 0.08 ) & 0.004  &  8.984  \\ 
 11.279687  &  41.621742  & vn.3.2.272 &  26.46  &  17.75 ( 0.03 ) & 17.99 ( 0.05 ) & 0.000  &  8.962  \\ 
 11.332550  &  41.788881  & vn.2.3.463 &  11.14  &  18.78 ( 0.03 ) & 19.10 ( 0.08 ) & 0.006  &  8.976  \\ 
 11.421142  &  41.747228  & vn.2.2.158 &  11.62  &  18.77 ( 0.03 ) & 19.11 ( 0.08 ) & 0.001  &  8.933  \\ 
 11.308658  &  41.764531  & vn.3.3.41 &  12.75  &  18.51 ( 0.03 ) & 18.70 ( 0.07 ) & 0.003  &  8.982  \\ 
 11.351142  &  41.735061  & vn.2.2.492 &  20.17  &  18.01 ( 0.03 ) & 18.31 ( 0.06 ) & 0.001  &  8.961  \\ 
 11.354621  &  41.708811  & vn.2.2.463 &  16.55  &  18.62 ( 0.03 ) & 19.01 ( 0.08 ) & 0.003  &  8.953  \\ 
 11.343588  &  41.903383  & vn.2.3.425 &  13.04  &  18.54 ( 0.03 ) & 18.89 ( 0.08 ) & 0.004  &  8.968  \\ 
 11.266917  &  41.933294  & vn.3.3.237 &  18.79  &  18.09 ( 0.03 ) & 18.60 ( 0.07 ) & 0.000  &  8.957  \\ 
 11.184892  &  41.927108  & vn.3.3.945 &  10.30  &  19.04 ( 0.03 ) & 19.26 ( 0.09 ) & 0.008  &  8.944  \\ 
 \tablebreak 
 11.331612  &  41.886650  & vn.2.3.468 &  22.15  &  18.05 ( 0.03 ) & 18.37 ( 0.06 ) & 0.002  &  8.972  \\ 
 11.154013  &  41.876450  & vn.4.3.114 &  11.47  &  19.07 ( 0.04 ) & 19.22 ( 0.09 ) & 0.001  &  8.962  \\ 
 11.163267  &  41.881942  & vn.4.3.54 &  13.88  &  18.68 ( 0.03 ) & 19.10 ( 0.08 ) & 0.002  &  8.961  \\ 
 11.143313  &  41.911761  & vn.4.3.181 &  13.54  &  18.34 ( 0.03 ) & 18.63 ( 0.07 ) & 0.000  &  8.941  \\ 
 11.345917  &  41.845825  & vn.2.3.413 &  16.38  &  18.35 ( 0.03 ) & 18.70 ( 0.07 ) & 0.002  &  8.974  \\ 
 11.282308  &  41.854561  & vn.3.3.173 &  18.92  &  18.26 ( 0.03 ) & 18.56 ( 0.06 ) & 0.002  &  8.982  \\ 
 11.303358  &  41.829725  & vn.3.3.77 &  12.83  &  18.76 ( 0.03 ) & 19.05 ( 0.08 ) & 0.002  &  8.983  \\ 
 11.307717  &  41.849969  & vn.3.3.54 &  35.91  &  17.24 ( 0.03 ) & 17.68 ( 0.04 ) & 0.001  &  8.980  \\ 
 11.231963  &  41.865203  & vn.3.3.731 &  11.18  &  18.71 ( 0.03 ) & 18.91 ( 0.08 ) & 0.004  &  8.979  \\ 
 11.098375  &  41.863444  & vn.4.3.420 &  14.66  &  18.30 ( 0.03 ) & 18.70 ( 0.07 ) & 0.002  &  8.953  \\ 
 11.111754  &  41.843253  & vn.4.3.353 &  20.23  &  17.92 ( 0.03 ) & 18.10 ( 0.05 ) & 0.002  &  8.967  \\ 
 11.445392  &  41.912014  & vn.2.3.69 &  14.64  &  18.29 ( 0.03 ) & 19.03 ( 0.08 ) & 0.001  &  8.949  \\ 
 11.410238  &  41.903642  & vn.2.3.203 &  10.97  &  19.05 ( 0.03 ) & 19.50 ( 0.10 ) & 0.004  &  8.958  \\ 
 11.415517  &  41.910472  & vn.2.3.178 &  11.15  &  18.79 ( 0.03 ) & 19.07 ( 0.08 ) & 0.001  &  8.956  \\ 
 11.455404  &  41.864358  & vn.2.3.21 &  16.12  &  18.54 ( 0.03 ) & 18.81 ( 0.07 ) & 0.002  &  8.943  \\ 
 11.380758  &  41.880753  & vn.2.3.314 &  10.43  &  18.98 ( 0.03 ) & 19.28 ( 0.09 ) & 0.006  &  8.965  \\ 
 11.381112  &  41.853336  & vn.2.3.311 &  31.61  &  17.44 ( 0.03 ) & 17.83 ( 0.04 ) & 0.001  &  8.965  \\ 
 11.367142  &  41.837386  & vn.2.3.354 &  24.20  &  17.67 ( 0.03 ) & 18.22 ( 0.05 ) & 0.003  &  8.969  \\ 
 11.375225  &  41.817983  & vn.2.3.333 &  17.29  &  18.71 ( 0.03 ) & 19.22 ( 0.09 ) & 0.003  &  8.965  \\ 
 11.394033  &  41.828908  & vn.2.3.260 &  14.93  &  18.11 ( 0.03 ) & 18.45 ( 0.06 ) & 0.002  &  8.960  \\ 
 11.393396  &  41.937289  & vn.2.3.265 &  15.34  &  18.28 ( 0.03 ) & 18.69 ( 0.07 ) & 0.004  &  8.957  \\ 
 11.243525  &  41.945469  & vn.3.3.325 &  28.61  &  17.45 ( 0.03 ) & 17.80 ( 0.04 ) & 0.001  &  8.949  \\ 
 11.576296  &  42.216006  & vn.1.4.92 &  10.97  &  19.29 ( 0.03 ) & 19.68 ( 0.11 ) & 0.000  &  8.884  \\ 
 11.575821  &  42.184472  & vn.1.4.93 &  12.33  &  18.79 ( 0.03 ) & 19.24 ( 0.09 ) & 0.000  &  8.893  \\ 
 11.450667  &  42.214064  & vn.2.4.28 &  21.15  &  18.20 ( 0.03 ) & 18.43 ( 0.06 ) & 0.000  &  8.869  \\ 
 11.464258  &  42.140753  & vn.2.4.2 &  10.22  &  18.69 ( 0.03 ) & 18.96 ( 0.08 ) & 0.000  &  8.901  \\ 
 11.169900  &  41.903150  & 091-2412 &  17.45  &  18.02 ( 0.03 ) & 18.43 ( 0.06 ) & 0.002  &  8.952  \\ 
 11.064200  &  41.569270  & vn.4.2.678 &  78.00  &  16.07 ( 0.03 ) & 16.50 ( 0.02 ) & 0.001  &  9.044  \\ 
 10.985150  &  41.217580  & vs.1.4.439 &  17.57  &  18.05 ( 0.03 ) & 18.38 ( 0.06 ) & 0.002  &  8.930  \\ 
 10.929470  &  41.247570  & D31.D.836 &  41.79  &  16.83 ( 0.03 ) & 17.30 ( 0.03 ) & 0.000  &  8.981  \\ 
 10.918150  &  41.185690  & vs.2.4.272 &  55.79  &  17.08 ( 0.03 ) & 17.52 ( 0.04 ) & 0.000  &  8.951  \\ 
 10.952310  &  41.153930  & vs.2.4.41 &  20.29  &  17.83 ( 0.03 ) & 18.12 ( 0.05 ) & 0.002  &  8.911  \\ 
 10.810760  &  41.504040  & 078-1587 &  21.67  &  18.01 ( 0.03 ) & 18.42 ( 0.06 ) & 0.004  &  9.054  \\ 
\enddata

\footnotetext{a}{$^a$Source of optical parameters for the Cepheids: 
vn*/vs* POMME \protect{\citep{fliri2012}}, 0* (Pan-STARRS); 
D* DIRECT \protect{\citep{2003AJ....126..175B}} }

 \footnotetext{b}{$^b$ includes absolute zeropoint uncertainty of 0.03 mag}
\end{deluxetable}

\tabletypesize{\small}
\begin{deluxetable}{cccccc}
\tablenum{2}
\tablecaption{PLCZ Fits to M31 Cepheids \label{table2}}
\tablehead{\colhead{Relation}&\colhead{Band}&\colhead{$\sigma$}&\colhead{$b$}
&\colhead{$\delta M/ \delta{\rm [O/H]}$} &\colhead{$m(\log P=1.2)$}}
\startdata
PL & $F160W$ &  0.174  &  -3.003 ( 0.127 ) &   & 18.292 ( 0.021 ) \\
PL & $F110W$ &  0.201  &  -2.725 ( 0.150 ) &   & 18.637 ( 0.024 ) \\
PLW & $F160W,F110W$ &  0.216  &  -3.432 ( 0.167 ) &   & 17.761 ( 0.026 ) \\
PLW & $F160W,F336W$ &  0.195  &  -3.433 ( 0.145 ) &   & 17.639 ( 0.024 ) \\
PLCZ & $F160W,F110W$ &  0.214  &  -3.419 ( 0.160 ) & -0.65 ( 0.73 )& 17.761 ( 0.026 ) \\
\enddata
\end{deluxetable}
\clearpage

The $F160W$ images for each pointing, consisting of 1600 s in 4 dithered 
exposures, were combined and resampled to a final scale of 0\farcsec08 pixel$^{-1}$. 
 The Cepheid photometry was measured by simultaneously fitting model PSFs to the
 Cepheid and any unresolved sources in its vicinity using the same zeropoint scale 
derived from the standard star P330E as in \citet{2011ApJ...730..119R}. 
Artificial stars were added to the images and measured to assess the crowding 
bias and to determine the photometric errors.  The mean bias was 0.002 mag and 
the mean statistical error was 0.01 mag.  With the $F110W$ filter, only a single 
dither of 700 or 800 s was obtained limiting the value of resampling the image 
on a finer scale and the use of PSF fitting, so we measured the $F110W$ flux in 
small apertures of 2 pixel radius. Cepheid parameters are given in Table~\ref{table1}.

\begin{figure}[ht]
\vspace*{140mm}
\figurenum{2}
\includegraphics{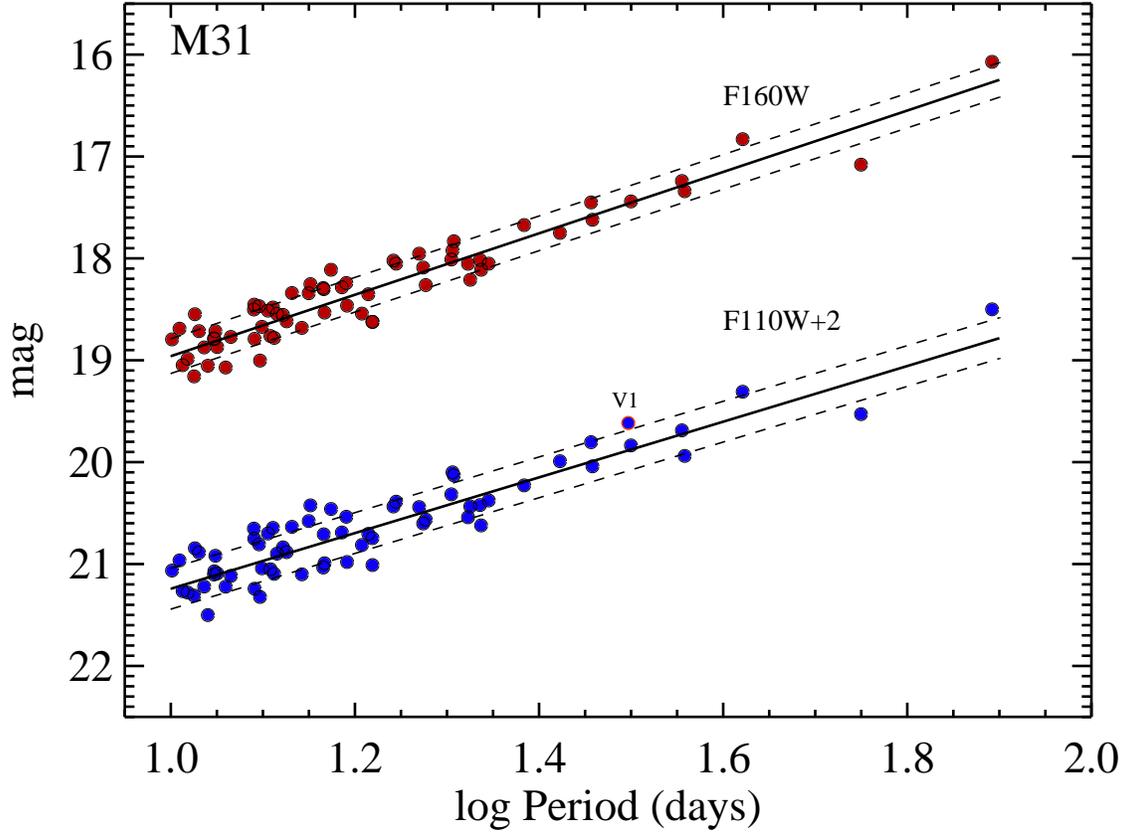}
\caption{Near-IR \PL relations for 69 Cepheids in M31 with 
$\log P > 1$ (Table~\ref{table1}).
The single slope fitted to the relations is given in Table~\ref{table2}, and
is shown as the solid lines. Dashed lines indicate the average
dispersion of 0.17 mag ($F160W$), a factor 3.5 smaller than 
previous ground-based optical \PL relations, and 0.20 mag 
($F110W$). \label{fig2}}
\end{figure}

The $F160W$ and $F110W$ \PL relations are shown in Figure~\ref{fig2} with relevant 
parameters in Table~\ref{table2}.  
 For historical interest, we have included one additional measurement in 
Figure~\ref{fig2}, the HST WFC3-IR $F110W$ observation of \citeauthor{1929ApJ....69..103H}'s
\citeyearpar{1929ApJ....69..103H}  first 
Cepheid discovered in M31, V1 with $P=31.4$ days, observed by his namesake 
telescope by the Hubble Treasury Program Team (P.I. K. Noll).  
The fitted slope of $-3.00 \pm 0.13$ for $F160W$ is 
in good agreement with the value of $-2.91 \pm 0.06$ found for 448 more distant
 Cepheids in 9 hosts with the same filter and instrument \citep{2011ApJ...730..119R}. 
 The M31 slope is about 1.5 $\sigma$  shallower than the slope of $-3.20 \pm 0.06$ 
 we estimate for the LMC Cepheids from \citet{2004AJ....128.2239P} 
after interpolating between the ground-based $J$- and $H$-band slopes.  The 
dispersion of the M31 Cepheids about this relation, 0.17 mag, is a factor of
 3.5 times lower than that of the optical \PL relations measured for M31 from 
the ground and even less than the 0.23 mag dispersion measured by 
\citet{2001ApJ...549..721M} 
with NICMOS, likely a result of the greater photometric stability of WFC3-IR.  
Although the dispersion is larger than the 0.13 mag dispersion measured by 
\citet{2004AJ....128.2239P} 
 in the $H$-band in the LMC, the difference is readily explained by our use 
of random phase measurements.  Resampling random phases from the 
\citet{2004AJ....128.2239P} 
light curves yields an average dispersion of 0.18 mag with no offset, a result 
nearly identical to ours, and confirming a similar intrinsic dispersion of 0.12 
mag.  A slope-insensitive distance indicator for our sample, the mean Cepheid 
magnitude at the sample mean period of $\log P = 1.2$, is $18.292 \pm 0.021$ mag, 
sufficient to measure the distance to M31 to 1\% given sufficient calibration.

The dispersion increases to 0.20 mag for $F110W$.  While some of the increase may 
result from additional differential extinction, most appears to come from the 
larger amplitudes of the light curves at shorter wavelengths.  The LMC Cepheids 
predict a random phase dispersion of 0.22 mag.  A strong correlation between the
 IR band residuals is apparent in Figure~\ref{fig2}, as expected from our
 use of random but coincident phases as well as from intrinsic variation.

Next we fit a Wesenheit relation (PLW in Table~\ref{table2}) to account for the effect of 
differential extinction along the inclined line of sight of the form 
$F160W-1.54(F110W-F160W)=a+b \log P$ where 1.54 is the value of $A_{F160W}$ per
 magnitude of $A_{F110W}-A_{F160W}$ for a 
\citet{1989ApJ...345..245C} 
reddening law with $R_V=3.1$.  The slope of this fit, $-3.43 \pm 0.17$, is 
in good agreement with the slope of $-3.38 \pm 0.09$  for the same relation 
in $J$ and $H$ for the LMC \citep{2004AJ....128.2239P}. 
A similar result was found using the $F336W$ UV color in place of $F110W$ with a reddening term of 0.14.  Next 
we included a metallicity parameter (PLCZ in Table~\ref{table2}) to account for any apparent
correlation between Cepheid near-infrared fluxes and the local value of $12+\log[O/H]$ measured from HII regions, a proxy 
for Cepheid metallicity.  Use of the deprojected radial gradient from
\citet{1994ApJ...420...87Z} 
results in an insignificant correlation of $-0.65 \pm 0.73$ mag per dex, 
compared to $-0.10 \pm 0.09$ mag per dex from the extragalactic Cepheids in 
\citet{2011ApJ...730..119R}. 
 The M31 Cepheids have little grasp on the Cepheid metallicity parameter because they lie in a 
narrow annulus along the disk \citep{fliri2012} with a full range of less than 0.2 dex
 ($12+\log[O/H]$=8.87 to 9.05).

To determine the distance to M31 as well as other parameters of interest we now 
make use of the sample of Cepheids measured in the near-infrared by  
\citet{2011ApJ...730..119R}, 
multiple anchors for the Cepheid distance scale, and SN data which can be used to 
determine the Hubble constant.  We follow the same formalism as in 
\citet{2011ApJ...730..119R}, 
solving one simultaneous system of linear equations which relate Cepheid magnitudes
 to their absolute magnitudes, slope of their \PL relations, local metallicity, 
distances to their hosts, and --with the inclusion of SN Ia data-- a measurement of 
the Hubble constant.  The Cepheids of M31 are assumed to share the same nuisance 
parameters as other Cepheids (i.e., luminosity, slope of \PL, metallicity 
dependence) but with a unique distance.  For the M31 Cepheids, the photometric 
system used to measure their colors was somewhat different.  While the Cepheids
 in the 8 SN Ia hosts, the maser host NGC 4258, and M31 were all measured with 
$F160W$ on WFC3-IR, the optical colors of the POMME M31 Cepheids, useful for dereddening, 
were not measured by Megacam or PHAT with the same $V$ and $I$ bands on HST as the others.  To 
account for this difference we employed one of two different prescriptions:  
1) we assumed a uniform value for the $V-I$ color of the M31 Cepheids as the 
mean of those measured from the ground by the DIRECT program, $V-I=1.23 \pm 0.03$, 
(indicated as fit $PLW=H_{V,I}$ in Table~\ref{table3}); or 2) we used the individual 
$F110W-F160W$ colors measured for the M31 Cepheids from the PHAT data with a small offset derived 
to give the same mean color correction in $V-I$ from 
DIRECT\footnote{By equating the mean $V-I$ dereddening with that for 
$F110W-F160W$ we can solve for a color offset to insure they have the same
 mean.  That is, $0.504<V-I> = 1.54<F110W-F160W-X>$, where $<V-I>$=1.23, gives 
$X=-0.066$ mag.} (indicated as $PLW=H_{J,H,X}$ in Table~\ref{table3}).   The 
advantage of the latter approach is that it can account for {\it differential} 
reddening along the line of sight while providing a reddening 
correction which is consistent with that used for non-M31 Cepheids.  We adopt an 0.03 mag systematic 
uncertainty for the use of colors measured with a different photometric system 
and an 0.04 mag systematic uncertainty between near-IR magnitudes of Cepheids 
measured on the ground and those measured from space.

\begin{deluxetable}{rccccccccccc}
\tablenum{3}
\rotate
\tablecaption{Global Fits \label{table3}}
\tablehead{\colhead{\#}&\colhead{$\chi^2_{dof}$}&\colhead{$\mu_{M31}$}
&\colhead{N}&\colhead{$H_0$}&\colhead{$<$P}
&\colhead{$\rm {[O/H]}$\tablenotemark{a}}
&\colhead{$\delta M/ \delta{\rm [O/H]}$} 
&\colhead{$b$}&\colhead{Scale}&\colhead{PLW}&\colhead{$R_V$}} 
\startdata
 1 & 0.92  & {\bf 24.415(0.052)} &      514 & \bf{73.60(2.35)} & Y & zkh & -0.09(0.09) &-3.21(0.03)& 4258+MW+LMC &   $H_{J,H,X}$ &  3.1 \\
 2 & 0.92  & 24.403(0.050) &      514 & 73.54(2.35) & Y & zkh & ------ &-3.21(0.03)& 4258+MW+LMC &   $H_{J,H,X}$ &  3.1 \\
 3 & 0.79  & 24.421(0.048) &      514 & 74.06(2.20) & Y & zkh & -0.13(0.09) &-3.19(0.02)& 4258+MW+LMC &   $H_{V,I}$ &  3.1 \\
 4 & 0.79  & 24.402(0.047) &      514 & 73.97(2.20) & Y & zkh & ------ &-3.19(0.02)& 4258+MW+LMC &   $H_{V,I}$ &  3.1 \\
 5 & 0.89  & 24.568(0.046) &      553 & 74.40(2.28) & Y & zkh & -0.12(0.09) &-3.08(0.02)& 4258+MW+LMC &   $H$ &  3.1 \\
 6 & 0.86  & 24.438(0.049) &      514 & 73.99(2.29) & Y & zkh & -0.10(0.09) &-3.19(0.03)& 4258+MW+LMC &   $H_{J,H,X}$ &  2.5 \\
 7 & 1.88  & 24.413(0.074) &      636 & 74.87(3.32) & Y & zkh & -0.09(0.13) &-3.17(0.04)& 4258+MW+LMC &   $H_{J,H,X}$ &  3.1 \\
 8 & 0.89  & 24.410(0.051) &      563 & 74.66(2.32) & N & zkh & -0.08(0.09) &-3.20(0.03)& 4258+MW+LMC &   $H_{J,H,X}$ &  3.1 \\
 9 & 0.92  & 24.400(0.051) &      514 & 73.98(2.46) & Y & T$_e$ & -0.09(0.13) &-3.21(0.03)& 4258+MW+LMC &   $H_{J,H,X}$ &  3.1 \\
10 & 0.84  & 24.468(0.070) &      448 & 73.78(3.11) & Y & zkh & -0.20(0.11) &-3.11(0.06)& 4258 &   $H_{J,H,X}$ &  3.1 \\
11 & 0.91  & 24.380(0.064) &      514 & 75.39(2.87) & Y & zkh & -0.18(0.12) &-3.20(0.03)& MW &   $H_{J,H,X}$ &  3.1 \\
12 & 0.90  & 24.500(0.098) &      514 & 71.41(3.41) & Y & zkh & -0.18(0.11) &-3.20(0.03)& LMC &   $H_{J,H,X}$ &  3.1 \\
13 & 0.92  & --- (---) &      514 & 74.00(2.25) & Y & zkh & -0.08(0.09) &-3.21(0.03)& 4258+MW+LMC+M31 &   $H_{J,H,X}$ &  3.1 \\
14 & 0.90  & --- (---) &      514 & 76.17(3.62) & Y & zkh & -0.18(0.11) &-3.20(0.03)& M31 &   $H_{J,H,X}$ &  3.1 \\
\enddata

\footnotetext{a}{$^a$ Note: metallicity calibration reference: 
zkh: \citet{1994ApJ...420...87Z}, T$_e$: \citet{2011ApJ...729...56B}.}
\end{deluxetable}
\clearpage

Our best distance estimate for M31 is $\mu_0=24.415 \pm 0.052$, a 2.4\% (statistical) distance 
determination 
which makes use of independent distance determinations to NGC
 4258, Milky Way Cepheid parallaxes, and DEB distances in the LMC 
\citep[see][for a description of these distance-scale anchors]{2011ApJ...730..119R}.  

The best fit parameters are given in Table~\ref{table3}, top line.  
 Column 2 gives the value of $\chi^2_{dof}$, Column 3 the distance modulus
of M31 and its statistical uncertainty, Column 4 the number of 
Cepheids used in the fit, Column 5 the value and total 
uncertainty in $H_0$,  Column 6 is a flag to indicate the use of
Cepheids below the optical completeness limit (see below), Column 7 gives
the metallicity calibration used, Column 8 the correlation coefficient in
the PLWZ regression, Column 9 the value and 
uncertainty of the slope of the Cepheid PLWZ relation. The next 
three columns are used 
to indicate variants in the analysis whose impact we now consider.

 Ignoring the metallicity parameter (lines 2 and 4), 
changing the method of color correction (line 3), 
changing the reddening parameter for the Cepheids from $R_V=3.1$ to 
$R_V=2.5$ (line 6), or 
changing the metallicity scale from \citet{1994ApJ...420...87Z} 
to that of \citet{2011ApJ...729...56B} (line 8), 
changes the distance to M31 by $< 0.02$ mag.  Failing to clip outliers in any 
Cepheid \PL relations (line 7) doubles the $\chi^2_{dof}$ but changes the distance to M31 
by $<$ 0.01 mag.  Retaining infrared measurements of 
Cepheids with periods below the optically-determined completeness limit 
(indicated by $<P=N$, line 8), has no effect on the distance.   
The only significant change in the distance to M31 occurs when discarding the 
color measurements used to account for extinction (line 5).  The 0.15 mag increase in 
the distance in this case indicates that the M31 Cepheids have {\it more} extinction, $\Delta A_H=0.15$ mag 
(or $Delta A_V=0.75$ mag)  than the average of the mostly 
extragalactic Cepheids in other hosts.  This is not surprising as M31 is more 
inclined than any of the other hosts.  We also used each of the 3 distance-scale 
anchors separately as shown in lines 10 to 12.  The uncertainties in the distances 
increase with the use of only a single anchor.  
The lowest uncertainty is from the use of 
the parallax measurements by \citet{2007AJ....133.1810B} 
to Milky Way Cepheids (line 11), resulting in $\mu_0=24.38 \pm 0.064$.  

Following the approach of \citet{2011ApJ...730..119R}, 
we quantify the {\it systematic} uncertainty in the distance to M31 from the dispersion 
of the variants in the analysis, $\sigma=0.03$ mag for the 12 variants not
 including the one neglecting color information which ignores the large extinction 
for the M31 Cepheids due to inclination.  Thus our best estimate of the distance
 to M31 is $\mu_0=24.42 \pm 0.05 \pm 0.03$ or 765 $\pm$ 28 kpc.  This 
distance is in excellent agreement with the frequently-cited measurement from 
\citet{1990ApJ...365..186F} 
 and sits near the middle of the range of past measurements summarized by 
\citet{2005MNRAS.356..979M}. 

Lastly, we made use of the two DEB measurements for M31 from 
\cite{2005ApJ...635L..37R} 
and \cite{2010A&A...509A..70V} 
with a mean of $\mu_0=24.36 \pm 0.08$ to determine the Hubble constant by 
their ability to calibrate the Cepheid luminosities.  Combined with the preceding three 
anchors (line 13), the use of the independent M31 distance has negligible impact, 
increasing the Hubble constant by 0.2 km s$^{-1}$ Mpc$^{-1}$ to 74.0 
km s$^{-1}$ Mpc$^{-1}$ with no reduction in uncertainty.  The present 
limitations of the use of 
M31 as an independent anchor are the lack of $V-I$ colors measured with HST WFC3 and 
the lower precision of its independent distance compared to the other 3
 anchors.  The use of M31 DEB results without any of the other anchors (line 14) 
yields a larger Hubble constant with larger error though still consistent with 
prior results. The best fit to the Hubble constant (line 1), which only makes use of the M31 Cepheids to constrain the slope and metallicity parameter, is 
73.6$\pm$2.4 km s$^{-1}$ Mpc$^{-1}$ and does not bring a significant 
difference to the one inferred by \cite{2011ApJ...730..119R}.

\section{Discussion}

The main advantages of the PHAT space-based observations of Cepheids 
presented here over previous data come from the reduction in extinction and crowding.  The
 result is the tightest \PL relation for M31 Cepheids yet seen, which, combined 
with past, external calibrations, yields the most precise distance measurement for M31.  
The improved precision over the past history of optical, ground-based \PL 
relations is striking and continues to indicate the value of this kind of data for 
measuring the Hubble constant and dark energy \citep{2011ApJ...730..119R}. 

To see more clearly the advantages of these spaced-based observations over 
those from the ground 
we simulated the effect of crowding for near-infrared Cepheid magnitudes 
obtained with good ground-based seeing
by measuring the flux contained in an aperture of radius $r$=0\farcsec9. 
 The dispersion increased from 0.20 to 0.24 mag in $F110W$, the slope of 
the \PL relation flattened by 0.5 and the intercept became brighter by 
0.3 mag.
This is not unexpected as the impact of crowding is greater for the 
lower-period, fainter Cepheids.  These results are similar to those 
determined for ground-based crowding in the optical by 
\cite{2000AJ....120..810M}. 
 It is important to note that the effect of crowding is far more severe for 
M31 due to its large inclination than for less inclined galaxies, even those 
much farther away.  In addition, Cepheids found from the ground via image 
subtraction are likely to suffer significantly {\it greater} crowding than those 
selected from PSF fitting as the addition of comparable constant flux to a 
PSF fit will reduce the amplitude of the light curve and remove the object 
from an amplitude-selected sample \citep{2000PASP..112..177F}.

The \PL relations for M31 may still improve in the near future.  Additional 
observations by the PHAT program should augment the Cepheid sample by dozens.  
Cepheids with $\log P < 1$, though less useful for distance scale work, can 
be mined from the data to study the short period end of the relation.  The 
use of Cepheid phase information from concurrent, ground-based optical 
monitoring of M31 can be used to recover the phase of the PHAT observations, 
reducing the scatter of the \PL by up to $\sim$ 50\%, an equivalent leverage 
as doubling the sample of measurements presented here.

\vfill

We are grateful to the members of the PHAT MCT Program led by Julianne Dalcanton and aided by Jason Kalirai for their tremendous
efforts to obtain the PHAT measurements and their support of this work.  
Financial support for this work was provided in part by the POMMME 
 project (ANR 09-BLAN-0228). 
This work is also based on observations obtained with MegaPrime/MegaCam, a joint 
project of CFHT and CEA/DAPNIA, at the CFHT which is operated by the National 
Research Council (NRC) of Canada, the Institut National des Sciences de l'Univers 
of the Centre National de la Recherche Scientifique (CNRS) of France, and the 
University of Hawaii.  JF acknowledges the hospitality of GEPI, where part of the POMME work was carried out.

\facility{\textit{Facilities}: HST, CFHT}

\bibliographystyle{apj}
\bibliography{m31}

\begin{thebibliography}{28}
\expandafter\ifx\csname natexlab\endcsname\relax\def\natexlab#1{#1}\fi

\bibitem[{{An} {et~al.}(2004){An}, {Evans}, {Hewett}, {Baillon}, {Calchi
  Novati}, {Carr}, {Cr{\'e}z{\'e}}, {Giraud-H{\'e}raud}, {Gould}, {Jetzer},
  {Kaplan}, {Kerins}, {Paulin-Henriksson}, {Smartt}, {Stalin}, \&
  {Tsapras}}]{2004MNRAS.351.1071A}
{An}, J.~H., {et~al.} 2004, \mnras, 351, 1071

\bibitem[{{Baade} \& {Swope}(1963)}]{1963AJ.....68..435B}
{Baade}, W., \& {Swope}, H.~H. 1963, \aj, 68, 435

\bibitem[{{Baade} \& {Swope}(1965)}]{1965AJ.....70..212B}
---. 1965, \aj, 70, 212

\bibitem[{{Benedict} {et~al.}(2007){Benedict}, {McArthur}, {Feast}, {Barnes},
  {Harrison}, {Patterson}, {Menzies}, {Bean}, \&
  {Freedman}}]{2007AJ....133.1810B}
{Benedict}, G.~F., {et~al.} 2007, \aj, 133, 1810

\bibitem[{{Bonanos} {et~al.}(2003){Bonanos}, {Stanek}, {Sasselov}, {Mochejska},
  {Macri}, \& {Kaluzny}}]{2003AJ....126..175B}
{Bonanos}, A.~Z., {Stanek}, K.~Z., {Sasselov}, D.~D., {Mochejska}, B.~J.,
  {Macri}, L.~M., \& {Kaluzny}, J. 2003, \aj, 126, 175

\bibitem[{{Bresolin}(2011)}]{2011ApJ...729...56B}
{Bresolin}, F. 2011, \apj, 729, 56

\bibitem[{{Brown} {et~al.}(2006){Brown}, {Smith}, {Guhathakurta}, {Rich},
  {Ferguson}, {Renzini}, {Sweigart}, \& {Kimble}}]{2006ApJ...636L..89B}
{Brown}, T.~M., {Smith}, E., {Guhathakurta}, P., {Rich}, R.~M., {Ferguson},
  H.~C., {Renzini}, A., {Sweigart}, A.~V., \& {Kimble}, R.~A. 2006, \apjl, 636,
  L89

\bibitem[{{Cardelli} {et~al.}(1989){Cardelli}, {Clayton}, \&
  {Mathis}}]{1989ApJ...345..245C}
{Cardelli}, J.~A., {Clayton}, G.~C., \& {Mathis}, J.~S. 1989, \apj, 345, 245

\bibitem[{Clerke(1903)}]{clerke1903problems}
Clerke, A. 1903, Problems in astrophysics (London: A. \& C. Black)

\bibitem[{{Dalcanton} \& {et al.}(2012)}]{dalcanton2012}
{Dalcanton}, J., \& {et al.} 2012, to be submitted

\bibitem[{{De Vaucouleurs} \& {Corwin}(1985)}]{1985ApJ...295..287D}
{De Vaucouleurs}, G., \& {Corwin}, Jr., H.~G. 1985, \apj, 295, 287

\bibitem[{{Ferrarese} {et~al.}(2000){Ferrarese}, {Silbermann}, {Mould},
  {Stetson}, {Saha}, {Freedman}, \& {Kennicutt}}]{2000PASP..112..177F}
{Ferrarese}, L., {Silbermann}, N.~A., {Mould}, J.~R., {Stetson}, P.~B., {Saha},
  A., {Freedman}, W.~L., \& {Kennicutt}, Jr., R.~C. 2000, \pasp, 112, 177

\bibitem[{{Fliri} {et~al.}(2006){Fliri}, {Riffeser}, {Seitz}, \&
  {Bender}}]{2006A&A...445..423F}
{Fliri}, J., {Riffeser}, A., {Seitz}, S., \& {Bender}, R. 2006, \aap, 445, 423

\bibitem[{{Fliri} {et~al.}(2012){Fliri}, {Valls-Gabaud}, \&
  {Magnier}}]{fliri2012}
{Fliri}, J., {Valls-Gabaud}, D., \& {Magnier}, E.~A. 2012, to be submitted

\bibitem[{{Freedman} \& {Madore}(1990)}]{1990ApJ...365..186F}
{Freedman}, W.~L., \& {Madore}, B.~F. 1990, \apj, 365, 186

\bibitem[{{Hubble}(1929)}]{1929ApJ....69..103H}
{Hubble}, E.~P. 1929, \apj, 69, 103

\bibitem[{{Lee} {et~al.}(2011){Lee}, {Riffeser}, {Koppenhoefer}, {Seitz},
  {Bender}, {Hopp}, {Goessl}, {Saglia}, {Snigula}, {Sweeney}, {Burgett},
  {Chambers}, {Grav}, {Heasley}, {Hodapp}, {Jedicke}, {Kaiser}, {Kudritzki},
  {Luppino}, {Lupton}, {Magnier}, {Monet}, {Morgan}, {Onaka}, {Price},
  {Stubbs}, {Tonry}, \& {Wainscoat}}]{2011arXiv1109.6320L}
{Lee}, C.-H., {et~al.} 2011, ArXiv e-prints http://arxiv.org/abs/1109.6320

\bibitem[{{Macri} {et~al.}(2001){Macri}, {Calzetti}, {Freedman}, {Gibson},
  {Graham}, {Huchra}, {Hughes}, {Madore}, {Mould}, {Persson}, \&
  {Stetson}}]{2001ApJ...549..721M}
{Macri}, L.~M., {et~al.} 2001, \apj, 549, 721

\bibitem[{{Madore} \& {Freedman}(1991)}]{1991PASP..103..933M}
{Madore}, B.~F., \& {Freedman}, W.~L. 1991, \pasp, 103, 933

\bibitem[{{Magnier} {et~al.}(1997){Magnier}, {Augusteijn}, {Prins}, {van
  Paradijs}, \& {Lewin}}]{1997A&AS..126..401M}
{Magnier}, E.~A., {Augusteijn}, T., {Prins}, S., {van Paradijs}, J., \&
  {Lewin}, W.~H.~G. 1997, \aaps, 126, 401

\bibitem[{{McConnachie} {et~al.}(2005){McConnachie}, {Irwin}, {Ferguson},
  {Ibata}, {Lewis}, \& {Tanvir}}]{2005MNRAS.356..979M}
{McConnachie}, A.~W., {Irwin}, M.~J., {Ferguson}, A.~M.~N., {Ibata}, R.~A.,
  {Lewis}, G.~F., \& {Tanvir}, N. 2005, \mnras, 356, 979

\bibitem[{{Mochejska} {et~al.}(2000){Mochejska}, {Macri}, {Sasselov}, \&
  {Stanek}}]{2000AJ....120..810M}
{Mochejska}, B.~J., {Macri}, L.~M., {Sasselov}, D.~D., \& {Stanek}, K.~Z. 2000,
  \aj, 120, 810

\bibitem[{{Persson} {et~al.}(2004){Persson}, {Madore}, {Krzemi{\'n}ski},
  {Freedman}, {Roth}, \& {Murphy}}]{2004AJ....128.2239P}
{Persson}, S.~E., {Madore}, B.~F., {Krzemi{\'n}ski}, W., {Freedman}, W.~L.,
  {Roth}, M., \& {Murphy}, D.~C. 2004, \aj, 128, 2239

\bibitem[{{Ribas} {et~al.}(2005){Ribas}, {Jordi}, {Vilardell}, {Fitzpatrick},
  {Hilditch}, \& {Guinan}}]{2005ApJ...635L..37R}
{Ribas}, I., {Jordi}, C., {Vilardell}, F., {Fitzpatrick}, E.~L., {Hilditch},
  R.~W., \& {Guinan}, E.~F. 2005, \apjl, 635, L37

\bibitem[{{Riess} {et~al.}(2011){Riess}, {Macri}, {Casertano}, {Lampeitl},
  {Ferguson}, {Filippenko}, {Jha}, {Li}, {Chornock}, \&
  {Silverman}}]{2011ApJ...730..119R}
{Riess}, A.~G., {et~al.} 2011, \apj, 730, 119

\bibitem[{{Vilardell} {et~al.}(2007){Vilardell}, {Jordi}, \&
  {Ribas}}]{2007A&A...473..847V}
{Vilardell}, F., {Jordi}, C., \& {Ribas}, I. 2007, \aap, 473, 847

\bibitem[{{Vilardell} {et~al.}(2010){Vilardell}, {Ribas}, {Jordi},
  {Fitzpatrick}, \& {Guinan}}]{2010A&A...509A..70V}
{Vilardell}, F., {Ribas}, I., {Jordi}, C., {Fitzpatrick}, E.~L., \& {Guinan},
  E.~F. 2010, \aap, 509, A70

\bibitem[{{Zaritsky} {et~al.}(1994){Zaritsky}, {Kennicutt}, \&
  {Huchra}}]{1994ApJ...420...87Z}
{Zaritsky}, D., {Kennicutt}, Jr., R.~C., \& {Huchra}, J.~P. 1994, \apj, 420, 87

\end{thebibliography}

\vfill
\eject

\end{document}